\documentclass{elsart}
\usepackage{amssymb}
\usepackage{epsfig}

\begin{document}

\newcommand \be  {\begin{equation}}
\newcommand \bea {\begin{eqnarray} \nonumber }
\newcommand \ee  {\end{equation}}
\newcommand \eea {\end{eqnarray}}

\begin{frontmatter}

\title{{\bf MICROSCOPIC MODELS FOR LONG RANGED VOLATILITY CORRELATIONS}}

\author[3]{Irene Giardina}
\author[1,2]{Jean-Philippe Bouchaud}
\author[4]{Marc M\'ezard}

\address[3]{Service de Physique Th\'eorique,
 Centre d'\'etudes de Saclay,  Orme des Merisiers, 
91191 Gif-sur-Yvette Cedex, France }
\address[1] {Service de Physique de l'\'Etat Condens\'e,
 Centre d'\'etudes de Saclay,  Orme des Merisiers, 
91191 Gif-sur-Yvette Cedex, France}
\address[2]{Science \& Finance, 109-111 rue Victor-Hugo, 92532 France}
\address[4]{Laboratoire de Physique Th\'eorique et Mod\`eles Statistiques
Universit\'e Paris Sud, Bat. 100, 91 405 Orsay Cedex, France}

\begin{abstract}
We propose a general interpretation for long-range correlation 
effects in the activity and volatility of financial markets. This 
interpretation is
based on the fact that the choice between `active' and `inactive' strategies 
is subordinated to random-walk like processes. We numerically demonstrate our 
scenario 
in the framework of simplified market models, such as the Minority Game model 
with an 
inactive strategy, or a more sophisticated version that includes some
price dynamics. We show that real market data can be surprisingly
well accounted for by these simple models.
\end{abstract}
\end{frontmatter}

\section{Introduction}

When looking at the time series of the 
daily returns in liquids markets, it appears to the naked eye
that the observed behaviour is markedly non Gaussian: fluctuations are intermittent 
and strong, localized outbursts of volatility can be identified.
This fact, known as {\it volatility clustering} 
\cite{volfluct2,MS,Book}, 
can be analyzed more quantitatively 
by looking at the daily volatility $\sigma_t$ (defined for
example, as  the average squared high frequency return). This quantity 
 follows a log-normal
distribution \cite{stanley}, and its temporal correlation function 
$\langle \sigma_t  \sigma_{t+\tau} \rangle$ can be 
fitted by an inverse power of the lag $\tau$, with a rather small exponent in
the range $0.1 - 0.3$ \cite{volfluct2,PCB,stanley,muzy}. 
This suggests that there is no characteristic time scale
for volatility fluctuations: outbursts of market activity can persist for 
rather short times (say a few days), but also for much longer times, months or even years. 
Besides, these 
{\it long ranged} volatility correlations are observed on many different 
financial markets and over different periods of time, with qualitatively similar
features: stocks, currencies, commodities or interest rates (see for example the 
{\sc bund} contract discussed in \cite{Book}). This suggests 
that a common mechanism is at the origin of this rather universal phenomenon. 
This universality, which is also observed for other  `stylized
facts' of financial markets (for example the power law behaviour
of  the return distribution) has attracted much attention, especially
among physicists. It indeed supports the idea of the market as a
complex interacting system of the kind usually studied in physics,
where complex behaviour arises from individual actions, not
crucially depending on the details of the microscopic interactions.

In this line of thought, the approach of agent based modelling has
been intensively pursued, and  important insights into
market dynamics have recently been gained
\cite{Bak,MG1,CB,Farmer,Stauffer,Hommes,Lux,hommes,Iori,farmerrev}. 
These models postulate some simple behaviour at the level of 
the agents and investigate  the resulting price dynamics. Their aim is 
to reproduce real market phenomenology and understand its distinctive
features in terms of microscopic mechanisms. Of course the possibilities 
of modelling are {\it a priori} numerous, since very little is known
about the relevant ingredients needed to construct a realistic artificial
market. In this
context two possible strategies of investigation can be 
followed. The first is to look for very simple models, which may be in 
some respects unrealistic, but where both analytical resolution and
intuitive understanding can be reached. Alternatively,  one can
focus on more complex models, trying to select the necessary
microscopic structure needed to reproduce the stylized facts, and  
identify the `universality class' of real markets.

In this paper we consider models of the two types and we 
look for the microscopic origin of volatility clustering.
We propose a simple and robust mechanism to account for the 
appearance of long-ranged correlations in the above simplified models
\cite{QFus}. 
We then argue that this  mechanism  also very naturally
operates in real financial markets, and accounts well for the empirical 
findings. Similar ideas (although quantitatively different) were recently 
discussed
in \cite{Huberman,Lux}. 

\section{Simple models: the Minority Game}

As an example of the first family of models we consider the Minority Game 
(MG) and its variants \cite{MG1,MG2,Johnson,random,CGGS,Moro}.
This model describes the behaviour of competing agents that can
choose between  different  individual strategies as a function of
their past  performance. It was originally introduced to model 
bounded rationality and inductive reasoning behaviour
\cite{arthur,MG1}, but it has since then become a paradigm for systems 
with heterogeneous adaptive agents. 

The setup of the model is very
simple. At a given time each agent $i$ can take two possible actions
$a_i=\pm 1$. Once that the global action $A(t)=\sum_i a_i$ is
determined, those agents who are in the {\it minority} group win whereas the
others lose. The game is repeated at each time step. It would of course be
trivial if the agents took their actions by tossing a coin. In fact it 
is not, since the opinion formation mechanism is not trivial at
all. Its distinctive properties are: 
\begin{itemize}
\item{Heterogeneities: }
Each  agent $i$ is endowed with $s=1,...,S$  ``strategies'' of action which are
randomly chosen among all possible 
strategies at the initial time, and are kept fixed throughout
the game.
\item{ Information: }
A public and common information $I$, (for example the m-steps past
history\cite{MG1}, or an exogenous signal \cite{random}) 
is available at each time step. Each agent processes this
information (the same for all) using one of his strategies, and decides of  
his action.
\item{Adaptivity: }
Agents evaluate the performance of their strategies. Agent $i$ gives
scores to each of his strategies according to its observed predictive power:
\be
P_{i,s}(t+1)=P_{i,s}(t)-a_{i,s} A(t) \ ,
\ee
where $s$ is a strategy label, and a simple minority payoff function
$-a A $ has been implemented. Then, at each time $t$ the  best
strategy, the one with the highest score,  is actually used to decide the action.
\end{itemize}
The collective behaviour of the MG is extremely rich and has been
extensively studied in numerical and analytical works. The main
feature is the possibility, by tuning the parameter $\alpha$ given 
by the ratio of the number of different `patterns' of information to the number
of agents, to obtain an `efficient' or an `inefficient' behaviour 
(in a sense very close to the analogous concept in economics). 
However, in its original version, this model is  rather remote from financial
markets; in particular, there is no price dynamics. Several attempts have 
been  made to generalize it and construct more realistic market models 
\cite{MG2,Johnson}.
As first noticed in \cite{Johnson}, if one also allows the agents to be
inactive,  intermittent volatility fluctuations can be generated. 
This fact confirms in the context of generalized MG what has been
observed also for real markets, that is that volatility fluctuations
are related to {\it activity} correlations \cite{Gopi,Bonanno}. 
For what concerns  our main point, i.e. to explain volatility
clustering,  this means that  we can focus on the temporal 
pattern of activity in these models. From this point of view, 
the simplest model which exhibits
long ranged activity correlations is the MG with inactive
strategies. Here we just consider the original setup of the MG, but we 
give to each agent one more `inactive' strategy whose prediction is
always $a_i=0$ (see also \cite{Canat,Johnson,marsili}). 
The number of active agents will then fluctuate from time to
time, and for this reason we shall refer to this version of the model
as Grand-Canonical MG.
As we shall see in the next sections, 
in this case it 
is rather intuitive to understand how the microscopic individual
patterns determine a specific behaviour of the activity correlations.
\section{More complicated market models}
In order to go beyond the above mentioned limits of the MG we have
studied a more complicated  market model 
that, while retaining the non trivial opinion formation structure of
the MG,  allows traders to switch between a `bond' market (representing the
inactive state) and a `stock'
market, and accounts  properly both  for their wealth  balance and for
market clearing. 
\\
As in the MG each agent has $S$ random active strategies plus an inactive
one. Besides he owns a number $\phi_i$ of stocks and $B_i$ of bonds
which he can trade at each time step. The price $X(t)$ of the stock
evolves in time according to a supply/demand dynamics which is defined 
through the following steps:
\begin{itemize}
\item{Information:}
As in the MG at each time $t$ an information is given to the
agents. We choose it to be given by the  $m$ last steps of the 
past history of the return time series:
$I(t)=(\chi(t-m),\cdots ,\chi(t-1))$ with $\chi(t)= {\mbox {sign}}
(X(t)-X(t-1))$.
In this sense, our traders are chartists on short time scales, and take
their decision based on the past pattern of price changes (note that this
decision is not the same for different agents).
\item{Opinion formation mechanism: }
As in the MG agents use  adaptive strategies to take a decision, which 
in this case can be `buy', `sell', or `do nothing'. 
Besides, each agent wishes to buy/sell 
a quantity proportional to his current  belongings,
that is $a_i=g B_i(t)/X(t)$ if he wants to buys, 
$a_i=-g \phi_i(t)$ if he wants to sell.
Also, each agent can act as a {\it fundamentalist} with a certain
probability $p_{fond}$ that depends on the difference between the observed 
price and some `reference' price that grows with the risk-free interest 
rate. This leads, on the long run, to a 
mean reverting behaviour of the price towards its reference value.
\item{Price formation mechanism: }
Once the individual offers $a_i(t)$ are made the price update is
defined by \cite{CB,Farmer}: 
\be
\frac{\delta X(t)}{X(t)}=\frac{X(t+1)-X(t)}{X(t)}= \lambda \sum_i a_i(t)  \ ,
\ee
where $\lambda$ is a measure of the liquidity of the market. 
Highly liquid markets correspond to small $\lambda$.
\item{Market clearing mechanism:}
We consider a two steps dynamics. First the
decisions/orders are made and the price is updated. Then 
a matching between supply/offer is realized, which may leave a certain 
amount of unfulfilled orders. More precisely,  the amount of stocks 
$\delta \phi_i^*$ {\it actually} traded by agent $i$ is $g_e {B_i(t)}/{X(t+1)}$
if he buys and $-g_e\phi_i(t)$ if he sells, where $g_e$ is such that $\sum_i
\delta \phi_i^* \equiv 0$. Unfulfilled orders are then 
removed from the order book. 
\item{Wealth dynamics: }
After this `market clearing' the individual wealths are updated
using the quantity $\delta \phi_i^*$ of stocks actually traded
by agent $i$, i.e.:
\be
\phi_i(t+1)=\phi_i(t) + \delta \phi_i^*\qquad B_i(t+1)=B_i(t)
(1+\rho)-\delta \phi_i^* X(t+1),
\ee
where $\rho$ represents the interest rate. The total wealth of agent $i$ is
therefore $w_i(t)=\phi_i(t) X(t)+B_i(t)$.
\item{Scores update:}
The scores of the strategies are finally updated according to their
real profitability over the interest rate $\rho$:
\be
P_{i,s}(t+1)=\mu P_{i,s}(t)-a_{i,s}(t-1)
\left[\frac{\delta X(t)}{X(t)} -\rho\right].
\ee 
When $\mu < 1$, only the recent past is used to assess the performance of the
strategies. 
\end{itemize}
The phenomenology of this model is very rich  and a detailed account of
our results will be published separately \cite{GBM}. From the previous 
description we see that there are at least four parameters which enter 
crucially in the model: $\rho, g, \lambda$ and one 
entering the precise definition of $p_{fond}$. Tuning
these parameters we observe two qualitatively different regimes:
\\
1) An {\it Oscillatory Regime} (for small values of $\lambda$ and $g$), where
speculative bubbles are formed, and finally collapse in sudden crashes induced
by the fundamentalist behaviour. In this regime, markets are not efficient, 
and a large fraction of the orders is (on average) unfulfilled.
\\
2) A {\it Turbulent regime} (for large $\lambda$ and $g$) where the `stylized' 
facts of liquid markets are well reproduced: the market is approximately 
efficient (although some persistent or anti-persistent correlations survive),
the returns follow a power law distribution, and volatility clustering is 
present. 

Both these two regimes present interesting features that can be analyzed.
However, for the objective of the present paper, we focus here on the
second one. In this regime both the volatility and the
activity show power-law like correlations (in time) which 
are quantitatively very similar to those observed in the much simpler
Grand-Canonical MG: see Figures 1 and 2 below. This strongly indicates that
volatility/activity correlations are uniquely determined by the
opinion formation structure which is common to the two models. In the
next section we propose an explanation of how this can happen.

\section{A simple universal mechanism}

In the above models, scores are attributed by agents to their 
possible strategies,
as a function of their past performance. In a region of the parameter space 
where these  models lead to an efficient market, the autocorrelation 
of the price 
increments is close to zero, which means that to a first 
approximation, no strategy can on average be profitable. This implies 
that the difference between the score of two strategies will locally behave, 
as a function of  time, as a random walk.
Now,  
the switch between two strategies occurs when their scores cross. Therefore,
in the case where each agent has two strategies, say one `active' (trading in 
the market) and one `inactive' (holding bonds), the survival time of any one of 
these strategies will be given by the return time of a random walk (the difference 
between the scores of the two strategies) to zero. The interesting point is that 
these return times are well known to be power-law distributed (see below):
this leads to the non trivial behaviour of the volume autocorrelation 
function. In other words, the very fact that agents compare the performance of 
two strategies on a random signal leads to a multi-time scale situation.
More formally, let us define the quantity $\theta_i(t)$ that is equal to $1$ 
if agent $i$
is active at time $t$, and $0$ if inactive. The total activity is given by 
$a(t)=\sum_i 
\theta_i(t)$. The time autocorrelation of the activity is given by:
\be
C_a(t,t')= \langle a(t) a(t') \rangle = \sum_{i,j} \langle 
\theta_i(t)\theta_j(t')
\rangle.
\ee
We will actually use in the following the so-called activity variogram, 
directly related to the autocorrelation through:
\be
V_a(t,t')=\langle \left[a(t)-a(t')\right]^2 \rangle= 
C_a(t,t)+C_a(t',t')-2C_a(t,t').
\ee
One can consider two extreme cases which lead to the same result, up to a 
multiplicative constant: (a) agents follow 
completely different strategies and have independent activity patterns, i.e.
$\langle \theta_i\theta_j\rangle \propto \delta_{i,j}$ or (b) agents follow 
very 
similar strategies, for example by all comparing the performance of stocks to 
that of bonds,
in which case $\theta_i=\theta_j$. In both cases, $C_a(t,t')$ is
proportional to $\langle \theta_i(t)\theta_i(t') \rangle$. This quantity can 
be 
computed in terms of the distribution $P(s)$ of the survival time $s$ of the 
strategies. More details can be found in \cite{QFus,GL}. 
When $P(s)$ has a finite first moment $\langle s \rangle$ 
(finite average 
lifetime of the strategies), then $C_a(t,t')$ is stationary, i.e. it 
only depends on the difference $\tau=t'-t$. Introducing the Laplace 
transforms $ \mathcal{L}C_a(E)$
and $ \mathcal{L}P(E)$ of $C_a(\tau)$ and $P(s)$, the general 
relation between the two quantities reads \cite{GL}:
\be\label{result}
E \mathcal{L}C_a(E)=\left(1-\frac{2[1- \mathcal{L}P(E)]}{\langle s \rangle E
[1+ \mathcal{L}P(E)]}
\right).
\ee
For an 
unconfined random
walk, the return time distribution decays as $s^{-3/2}$ for large $s$ 
and therefore its first moment is infinite. However, 
in all the models mentioned above, there exist `restoring' forces 
which effectively confine the scores to a finite interval \cite{GBM}. This 
can be attributed, 
both in 
the case of the  MG or of more realistic market models, to `market 
impact', which
means that good strategies tend to deteriorate because of their very use,
or to the finite memory time of market operators used to
assess their strategies (cf. the parameter $\mu$ defined above). 
The consequence of
these effects is to truncate the $s^{-3/2}$ tail for values of $s$ larger 
than a certain 
equilibrium time $s_0$. Therefore, the first moment of $P(s)$ actually 
exists, such that 
Eq. (\ref{result}) is valid. Nevertheless, one can see
from Eq. (\ref{result}) that the characteristic 
$s^{-3/2}$ behaviour of $P(s)$ for short time scales leads to $E\mathcal{L}C_a(E) 
\sim
1 - B/\sqrt{E} + ...$ for $s_0^{-1} \ll E \ll 1$. This in turn leads to a 
singular behaviour
for the variogram $V_a(\tau)$ at small $\tau$'s, as $V_a(\tau) \propto 
\sqrt{\tau}$, before saturating to a finite value for $\tau \sim
s_0$. Intuitively, this  means that the probability for the activity to have 
changed significantly between $t$ and $t+\tau$ is
proportional to $\int_0^\tau ds \ s P(s) \propto \sqrt{\tau}$ (for $\tau \ll 
s_0$), where $s P(s)$ is the probability to be at time $t$ playing a
strategy with  lifetime $s$.

\begin{figure}
\epsfig{file=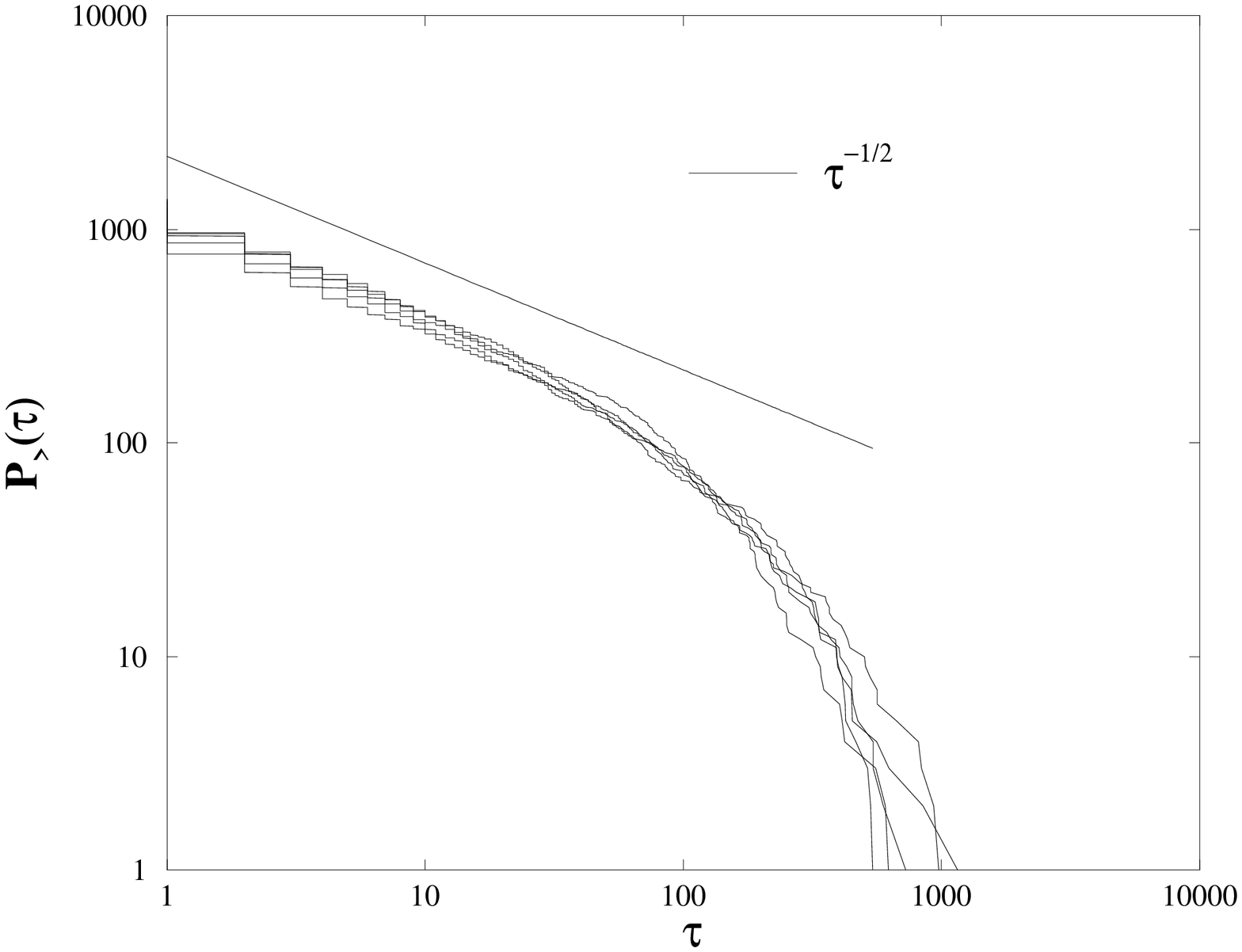,width=5cm,angle=270}\hspace{1cm}\epsfig{file=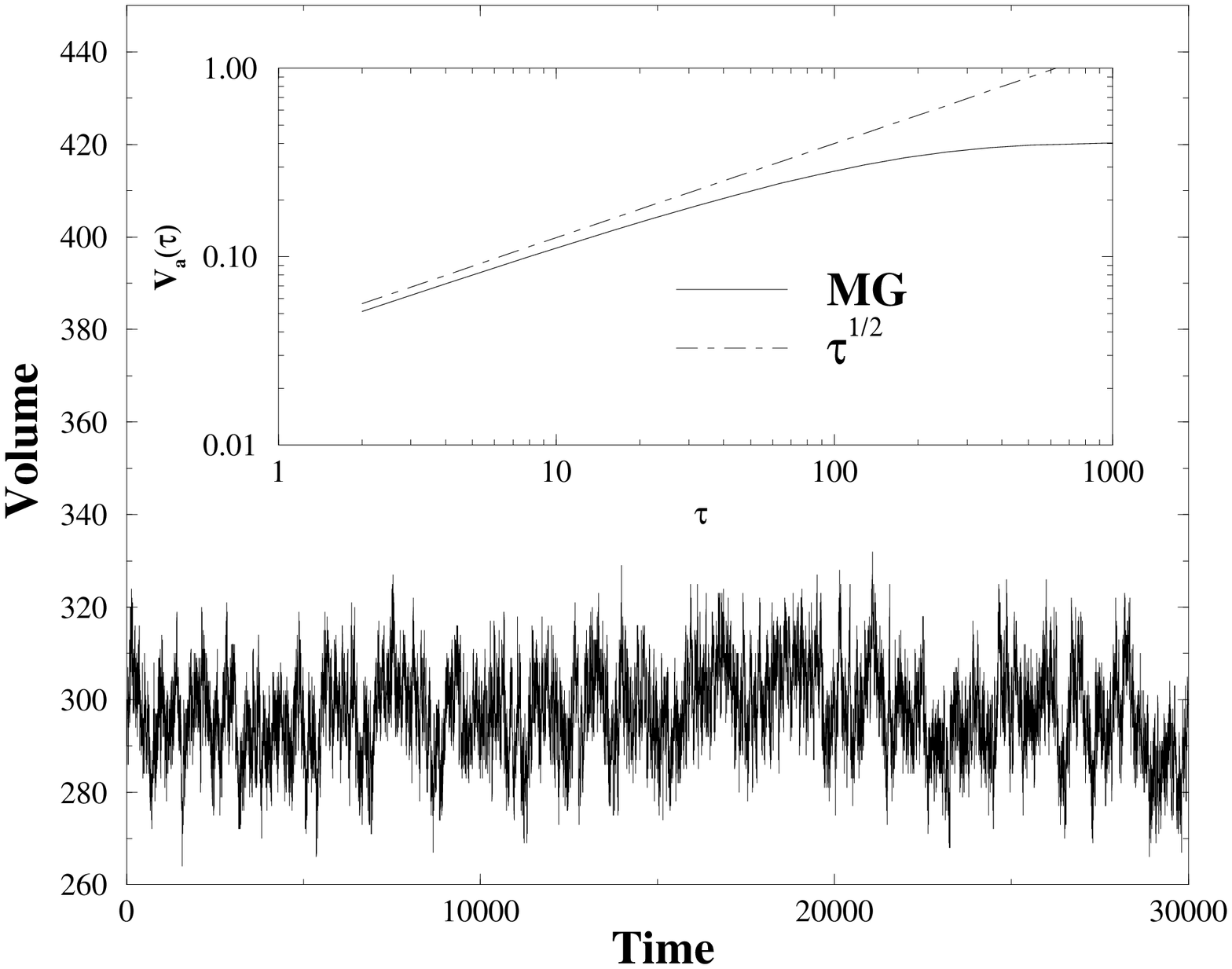,width=5cm,angle=270}
\vskip 0.3cm \caption{\small 1-a: Cumulative distributions for the survival
times of the active strategy in the Grand-Canonical MG with an
inactive  strategy and two active strategies, for 
five different agents. The parameter $\alpha$ is set to $\alpha=0.51$ 
($\alpha_c \simeq 1.$), and the number of agents is $501$. 
1-b: Volume of activity (number of active agents) as 
a function of time for the
MG with the same values of parameters.
Inset: The corresponding  activity variogram 
as a function of the lag $\tau$, in a log-log plot to emphasize the 
$\protect\sqrt{\tau}$ singularity at small $\tau$'s.
\label{fig3} }
\end{figure}

Let us illustrate and justify this general scenario 
for  the  MG with an inactive strategy. 
In this model  \cite{Canat,GBM} there is 
a critical value $\alpha_c$ of the relevant parameter 
 above which all agents finally become 
inactive. However, below this value, the system is always
`efficient' and our assumption about the local random-walk behaviour 
of the strategy scores is thus well grounded. However, in order to
confirm this point we show in Figure 1-a the cumulative distribution of the survival
time of the active strategy, for five different
agents. In the case of a pure random-walk this distribution should
behave as $s^{-1/2}$. We can see that this is precisely what happens
up to a certain time scale $s_0$ (which does not fluctuate 
much from agent to agent) above which the distributions are truncated. 
In Figure 1-b, for a value of $\alpha$ smaller 
than $\alpha_c$, we have plotted the volume of activity as a function of the time.
In the inset, the activity variogram $V_a(\tau)$ reveals
the characteristic $\sqrt{\tau}$ singularity discussed above, before 
saturating for
large $\tau$ ($\sim s_0$). 
This $\sqrt{\tau}$ singularity is present in the whole active phase 
$\alpha < \alpha_c$; $s_0$ is found to be proportional to $N$ (for $\mu=1$). 
The  analogous variograms for the volume of activity and for the
volatility  of the more realistic market model of section 3 are shown 
in Figure 2, together with the result of the MG. The two
behaviors are very similar and confirm  the universality of this
result \cite{GBM}. Note that the activity variogram of the MG model
can be very accurately fitted by the following simple form:
\be\label{MGfit}
\left.V_a(\tau)\right|_{fit}= V_a^\infty \left(1 -\exp(-\sqrt{\frac{\tau}{\tau_0}})\right).
\ee
We note that a similar mechanism might also be present in other
models where agents can switch between different strategies or classes
of strategies. For example, in \cite{Lux} each agent can  behave 
either as a fundamentalist or as a  trend follower, 
switching  between the two strategies as a function of their relative 
performance. For this model it has been observed that the activity bursts 
are indeed associated to a large number of agents switching 
from one behaviour to the other. 
The importance of the fact that these strategies are on average 
equivalent was also clearly stressed \cite{Lux} (see also \cite{Huberman}). 
It would be interesting to see if the 
activity variogram in this model can also be fitted using Eq. (\ref{MGfit}).
Another example is the adaptive
evolutionary system of \cite{hommes}, where agents can use different
expectation functions to estimate future prices and dividends (and then act
differently according to a mean-variance optimization procedure) and
choose between them according to a evolutionary performance measure. 
\footnote{In this model the presence of positive dividends on stocks is crucial
to implement a reasonable market dynamics. In the market model described above, 
on the other hand, we do not expect dividends to change substantially the dynamics
since a dividend at time $t$ is in fact followed by an immediate decrease of the price
$X(t)$ of the same amount, leaving the scores updates unchanged.}

\begin{figure}
\begin{center}
\epsfig{file=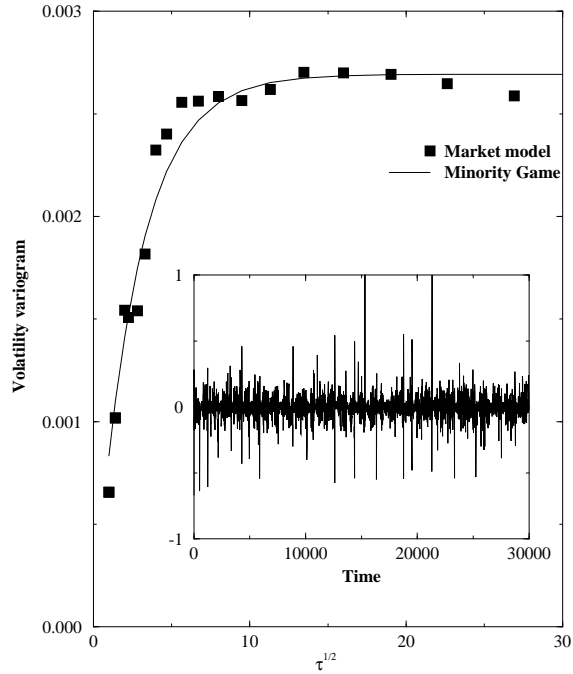,width=6cm}
\vskip 0.3cm \caption{\small 
Returns (relative price differences) as 
a function of time in the market model described in the text (section 3).
Large frame: The corresponding volatility variogram 
as a function of the lag $\tau$, in a log-log plot to emphasize the 
$\protect\sqrt{\tau}$ singularity at small $\tau$'s. The full line is
the fit Eq. (\protect\ref{MGfit})
\label{figm} }
\end{center}
\end{figure}
It is interesting to compare the above results with real market data. Figure 
3 shows the 
volume of activity on the S\&P 500 futures contract in the years 1985-1998. 
On the same graph, we have
reproduced the MG fit Eq. (\ref{MGfit}). Both the time scale and the volume scale
(arbitrary in the  MG model) have been adjusted to get the best agreement.
Furthermore, a constant has been added to $V_a$ (corresponding to a 
$\delta_{\tau,0}$ contribution to $C_a$), to account for the fact that part 
of  the trading activity is
certainly white noise (e.g. motivated by news, or by other non strategic causes). 
As can be seen, the agreement is rather good. Most significant is the 
clear $\sqrt{\tau}$ behaviour at small $\tau$ ($\tau < 50$ days). 
\begin{figure}
\begin{center}
\epsfig{file=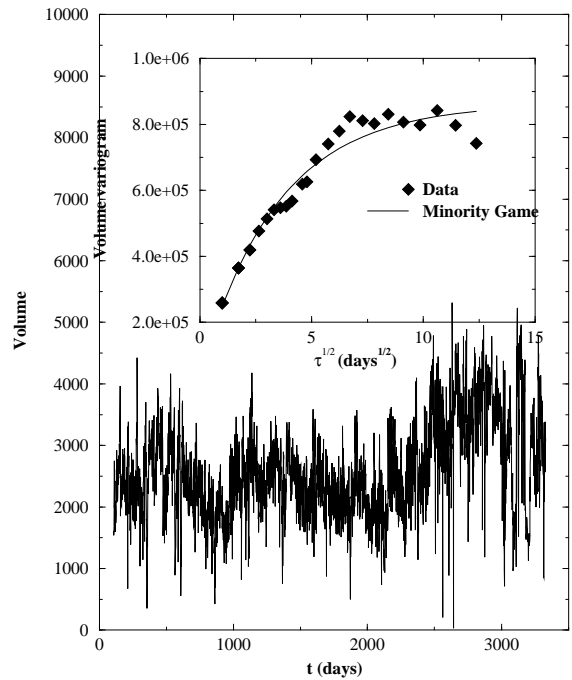,width=6 cm}
\vskip 0.3cm \caption{\small Total daily volume of activity (number of trades) on 
the S\&P 500 futures contracts in the years 1985-1998: compare to Figure 1 and 2. 
Inset: 
Corresponding variogram 
(diamonds) 
as a function
of the square-root of the lag. Note the clear linear behaviour for small $
\protect\sqrt{\tau}$. The full line is the MG fit Eq. (\protect\ref{MGfit}), with both 
axis
rescaled and a constant added to account for the presence of `white noise' 
trading.
\label{fig4} }
\end{center}
\end{figure}
We therefore
suggest that the effect captured by the  MG  or more 
sophisticated variants (Figures 1 and 2), 
namely the subordination of the activity on random walk like signals, is also 
present in real markets. 
It seems to us that this makes perfect sense since market participants indeed 
compare the results of different strategies to decide whether they should
remain active in a market or leave it. Note furthermore that although 
our scenario is based on the comparison between the scores
of strategies, similar results would be obtained if the volume was subordinated 
to the difference between the price and certain `psychological
levels' (i.e. the value 1000 for the S\&P, etc.). It is indeed 
reasonable that such 
levels play a role in determining the activity on financial markets
(see however \cite{psyco} for a discussion of this point).

\begin{figure}
\hspace*{+2cm}
\epsfig{file=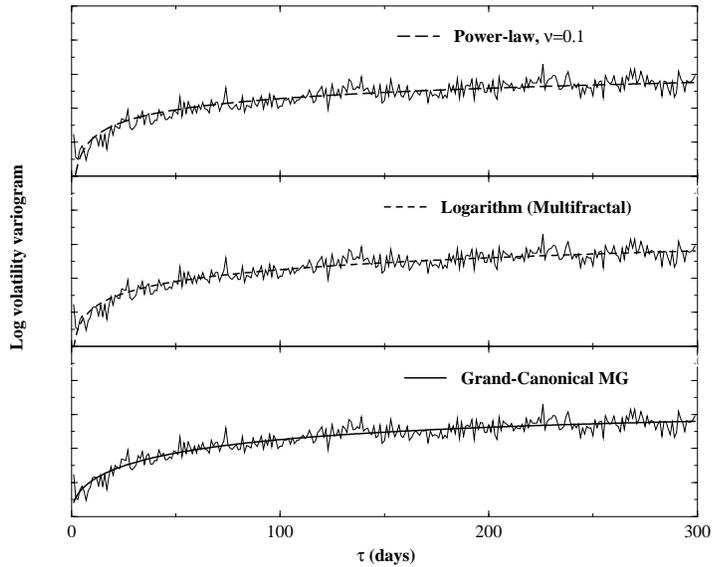,width=8cm,angle=270}
\vskip 0.3cm \caption{\small Variogram of the log-volatility, $\langle
\log^2(\sigma_t/\sigma_{t+\tau})\rangle$ as a function of $\tau$, averaged over 
17 different 
stock indices (American, European, Asian). The upper panel shows a fit with a
power law of time, the middle panel a fit with the logarithmic (multi-fractal) prediction, whereas the lower panel shows the MG fit Eq. (\protect\ref{MGfit}), with again both 
axis
rescaled and a constant added to account for the presence of `white noise' 
trading. Note that the three models lead to nearly indistinguishable fits.
\label{fig5}}
\end{figure}
Since the volatility and the volume of activity are
strongly correlated also in financial markets \cite{Bonanno,Gopi}, our 
interpretation should carry over to volatility fluctuations 
as well. This is illustrated in Figure 4 (lower panel), where the variogram of the log-volatility for major stock indices is shown, together with the very same
 MG fit. Again, the agreement is very good. 
We have however also shown for comparison the prediction of the multi-fractal
model of ref. \cite{muzy}  $V_a(\tau)=2\lambda^2 \log(\tau/\tau_0)$
(middle panel). We note that  the two models, although very different, lead to nearly 
indistinguishable numerical fits over the time scale considered. 
In the literature, it is also
customary to fit the volatility correlations with power laws. We have 
therefore also shown in the upper panel the corresponding fit for the variogram
$V_a(\tau)=c_0-c_1/\tau^{\nu}$, with $\nu=0.1$. This fit is also consistent with the
data (and is numerically very close to a logarithmic function because $\nu$ is small).
However, neither the power-law nor the logarithm have clear microscopic motivations. 
If, as we
believe, the simple mechanism that we have advocated is at the origin
of the correlations behaviour then the truly universal feature is
in the $\sqrt{\tau}$ behaviour of the
variogram for small $\tau$,  and {\it not} in the long 
time behaviour of the correlation function. In other terms, the exponent $\nu$ might 
be an effective, non universal exponent masking the true universal $1/2$ exponent 
that describes the initial increase of the variogram.

In summary, we have proposed a very general interpretation for long-range 
correlation 
effects in the activity and volatility of financial markets. This 
interpretation is
based on the fact that the choice between different strategies is 
subordinated to random-walk 
like processes. We have numerically demonstrated our scenario in the 
framework of simplified 
market models, and showed that, somewhat surprisingly, real market data can 
actually be quite accurately accounted for by these simple models (see Figs 3 
and 4). 

{\it Acknowledgments} 
Useful interactions with G. Canat, A. Cavagna, D. Challet, 
D. Farmer, A. Matacz, E. Moro, J.P. Garrahan, N.  Johnson, T. Lux, M.
Marsili,  M. Potters, P. Seager, D. Sherrington and H. Zytnicki are acknowledged.

\end{document}